\documentclass[12pt,a4paper]{article}

\title{Eppur, si muove !}

\author{Ll.\ Bel\\
\emph{Fisika Teorikoa, Euskal Herriko Unibertsitatea}, \\
\emph{P.K. 644, 48080 Bilbo, Spain} \\
\emph{e-mail: wtpbedil@lg.ehu.es}
}

\begin{document}
\maketitle

\begin{abstract}

Simple ideas that shed new light on the physics of rotation as it
concerns two famous experiments: The Wilson and Wilson, and the
Michelson and Morley experiments.

\end{abstract}

\section*{Introduction}

Yes indeed! The Earth rotates, the Moon and the Planets rotate, and
rotation is an ubiquitous state of motion in the Universe. But also in
laboratory physics rotation is often an essential component in some
experiments, among them the Michelson and Morley experiment, the Wilson and Wilson, 
and some others (\cite{Then}) that we shall not consider in this paper. 
Despite of this, many fundamental aspects of the physics of
rotation remain not well understood if Special relativity is needed, and
in particular when rotation and electromagnetism both play a role in an
experiment.

The very definition of a rotating frame of reference in Special
relativity has been questioned in the past and still is by some 
(\cite{Trocheris}-\cite{Herrera}). It
puzzles some authors the fact that the usual definition looks like non
relativistic and leads to time-like congruences that do not fill the
whole Minkowski's space. Many authors also are tempted to deal with
rotation invoking the Principle of local Lorentz invariance but this
amounts to overlooking some subtle points and the conditions that would
justify this simplification need to be clarified (\cite{Mashhoon1}-\cite{Mashhoon2}). 
Also is not always clear
whether it makes a difference to distinguish a rigid body that rotates from
incoherent dust that rotates as if it were rigid. These and other
thoughts recurrently come to the mind. 

In Section 1 we review and develop a decomposition of the
Lorentz transformations as a product of three transformations with
different meanings and different importance. This is to clarify what is
essential and what is just convenient in dealing with a particular
problem involving only Galilean frames of reference. 

In Section 2 we extend the point of view of the preceding section to
deal with the coordinate transformations between adapted coordinates from
a Galilean frame of reference to a rotating one and viceversa. The deep
novelty is that to be able to have the analog of the third factor in the
decomposition to select Cartesian-like coordinates of space, we feel
that a departure of common ideas about the concept of space in a
rotating frame of reference is necessary, as it was in some other problems 
(\cite{Bel96}-\cite{Bel02}). This has implications on the
Principle of local Lorentz invariance.

Section 3 is our late contribution to a lively recent polemic about the
outcome and the theory of the Wilson and Wilson experiment 
(\cite{Wilson}-\cite{Hertzberg}). 
Second order relativistic effects are completely negligible in the conditions of the
experiment and therefore usually only the first order result that agrees with the
experiment is calculated. We derive here an exact formula as a test case
of the procedure that we propose that is complete and remains simple.

As it is well known the Michelson and Morley experiment (\cite{MM}) has
been usually understood as a test of Special relativity in its most
restrictive sense (\cite{Brillet}-\cite{Lipa}), but not always
(\cite{Aspden}-\cite{Klauber2}). The usual attitude excludes taking into account the
rotation of the Earth except for changing the orientation of the
interferometer with respect to an hypothetic absolute space whose
existence would violate Special relativity. We have opposed this point
of view in three preceding papers (\cite{ABMM}-\cite{BelMolina})
predicting a positive result which by no means would contradict Special
relativity, but only some restrictive ways of dealing with this theory.
In these papers, besides the rotation of the Earth, also its gravity and
its oblateness, or the influence of the Sun and the Moon, were part of
the discussion and this somehow obscured a little the principal role of
the first. In Section 4 we summarize a bare-bones presentation of our
point of view emphasizing the role of the rotation of the Earth. 
\section{Galilean frames of reference}
Let us consider Minkowski's space-time and a Cartesian
system of coordinates $T$ and $X^i$ such that the line 
element is:

\begin{equation}
\label {1.1}
ds^2=-dT^2+\delta_{ij}dX^idX^j, \quad c=1, \quad i,j,k,..=1,2,3
\end{equation} 
This simple formula hides the rich geometrical structure of a
Galilean frame of reference, say $S_0$, which is a structure with two
ingredients that it is useful to exhibit explicitly.

i) The first ingredient is the congruence ${\cal C}_0$ of time-like
world-lines with parametric equations:

\begin{equation}
\label {1.2}
T=T, \quad X^i=const.
\end{equation}   
It is a Killing congruence, and therefore the motion that it defines is
rigid: 

\begin{equation}
\label {1.5}
\frac{d }{dT}\int_{L_g}ds=0,
\end{equation}
where $L_g$ is any segment of geodesic whose end-points lie on any two
world-lines of ${\cal C}_0$ and is orthogonal to them.

ii) The second ingredient is the foliation with equation:

\begin{equation}
\label {1.3}
T=const.
\end{equation}
It is a family of hyperplanes orthogonal to the congruence ${\cal C}_0$. The associated
synchronization is the scale that to an interval of $T$ along
any of the wold-lines, say $L_0$, corresponds the proper time interval:

\begin{equation}
\label {1.4}
T_1-T_0=\int_{L_0}\frac{ds}{dT}\,dT 
\end{equation}   

The purpose of this paper is twofold. First of all to unravel the
meaning and the role of the two ingredients in the relationship between
two different Galilean frames of reference. And secondly to apply our
findings to reach a better comprehension of uniformly rotating frames of
reference which will be the subject of the remaining of this paper
starting in Sect. 2. 

Let us consider the following formulae:

\begin{equation}
\label {1.6}
T=T, \quad X^i=x^i+v^iT
\end{equation}
and the inverse:

\begin{equation}
\label {1.6.1}
T=T, \quad x^i=X^i-v^iT
\end{equation}
where $v^i$ are constants. They make sense both in Classical mechanics
and in Special relativity and in both cases they have the same dual
meaning. Here we consider them in the framework of the Galilean frame of
reference considered above, and therefore we assume that: 

\begin{equation}
\label {1.7}
v^2<1, \quad v^2=v_iv^i, \quad v_i=\delta_{ij}v^j
\end{equation}

Formulae (\ref{1.6}) can be interpreted as the parametric equations
$X^\alpha(T;x^i)$ ($\alpha, \beta, \gamma\,..=0,1,2,3$) of a family of time-like world-lines with parameter $T$
and initial conditions $x^i$. This family is, as it was the case with ${\cal
C}_0$, a new Killing congruence ${\cal C}_1$ of time-like geodesics. They can
therefore be considered as the first ingredient of a new Galilean
frame of reference, say $S_1$.   

But the same formulae can be interpreted as defining a coordinate
transformation $X^\alpha(T, x^i)$ from coordinates $T,x^i$ to coordinates
$T,X^i$. From this point of view they define a coordinate
transformation between two Galilean frames of reference whose first
ingredient are two different congruences ${\cal C}_0$ and ${\cal C}_1$
but share the same foliation. This does not pose any problem but one has to keep in mind 
that the synchronization which was associated with proper-time along the
geodesics of ${\cal C}_0$ does not correspond now to proper-time along
${\cal C}_1$. This asymmetry may be inconvenient in some occasions and
convenient in some others. More on that later.

Using the coordinate transformation (\ref{1.6}) the line-element (\ref{1.1})
becomes:

\begin{equation}
\label {1.8.0}
ds^2=-(1-v^2)dT^2+2v_idx^idT+\delta_{ij}dx^idx^j
\end{equation}
that can be decomposed as follows:

\begin{equation}
\label {1.8}
ds^2=-{\theta^0}^2+ds_1^2
\end{equation}
where:

\begin{eqnarray}
\label {1.9a}
\theta^0&=&-\sqrt{1-v^2}dT+\frac{v_idx^i}{\sqrt{1-v^2}} \\
\label {1.9b}
ds_1^2&=&(\delta_{ij}+\frac{v_iv_j}{1-v^2})dx^idx^j
\end{eqnarray}

Let us now choose as a new foliation the family of hyperplanes
orthogonal to ${\cal C}_1$ and as an associated synchronization that
corresponding to choosing as scale of time the proper-time along any
of the world-lines of this congruence, say for instance:

\begin{equation}
\label {1.11}
L_0\ :\ X^i=x^i_0+v^iT
\end{equation}
This amounts to choosing the new time coordinate $t$ such that:

\begin{equation}
\label {1.10}
t-t_0=\int^{T_0}_0
\sqrt{1-\delta_{ij}\frac{dX^i}{dT}\frac{dX^j}{dT}}\,dT=\sqrt{1-v^2}T_0
\end{equation}
$t_0$ being an arbitrary origin of $t$ and $T_0(T,X^i;x^i_0)$ being the value
of $T$ at the intersection of $L_0$
with the hyperplane $\Pi$ of the new foliation passing through the event with
coordinates $T, X^i$.

From the equation of $\Pi$:

\begin{equation}
\label {1.12}
(X^i-X^i_0)v_i-(T-T_0)=0
\end{equation}
and from:

\begin{equation}
\label {1.12.1}
X^i_0=x^i_0+v^iT_0
\end{equation}
we obtain:

\begin{equation}
\label {1.13}
t-t_0=\frac{T-(X^i-x^i_0)v_i}{\sqrt{1-v^2}}
\end{equation}
and also, using (\ref{1.6}):

\begin{equation}
\label {1.14}
T=\frac{t-t_0}{\sqrt{1-v^2}}+\frac{(x^i-x^i_0)v_i}{1-v^2}
\end{equation}

Substituting $T$ in (\ref{1.8.0}) we get:

\begin{equation}
\label {1.15}
ds^2=-dt^2+ds_1^2
\end{equation} 
with $ds_1^2$ being the unchanged metric given in  (\ref{1.9b}).
This metric is obviously flat because its coefficients are constants
but the coordinates $x^i$ are not Cartesian coordinates. Particular Cartesian
ones are :

\begin{equation}
\label {1.16}
{\bar x}^i=(\delta^i_j+bv^iv_j)x^j, \quad
b=\frac{1}{v^2}(-1+\frac{1}{\sqrt{1-v^2}})
\end{equation}
as can be seen substituting the inverse functions:

\begin{equation}
\label {1.17}
x^i=(\delta^i_j+av^iv_j){\bar x}^j, \quad
a=\frac{1}{v^2}(-1+{\sqrt{1-v^2}})
\end{equation}
into (\ref{1.9b}) that becomes:

\begin{equation}
\label {1.18}
ds_1^2=\delta_{ij}d{\bar x}^id{\bar x^j}
\end{equation}
At the end we obtain thus:

\begin{equation}
\label {1.19}
ds^2=-dt^2+\delta_{ij}d{\bar x}^id{\bar x^j}
\end{equation}

We have therefore succeeded in decomposing a Lorentz transformation
leading from (\ref{1.1}) to (\ref{1.19}) in
three steps as
a product of three particular transformations with different meanings 
and different importance.

i) Step 1.- The first transformation is given by (\ref{1.6}) and it is
the one that picks the congruence, i.e. the rigid motion of a second
Galilean frame of reference. Any other choice would not define a new
frame of reference with uniform constant velocity. We shall see that this is
the only step of a Lorentz transformation that is necessary to
transform tensor fields between two different Galilean frames of
reference.

ii) Step 2.- The second transformation is the time transformation
given by (\ref{1.14}). This defines a convenient foliation and it is important
because the corresponding synchronization 
implements a fundamental
postulate of Special, as well as General, relativity identifying
proper space-time intervals along time-like world-lines with physical 
time intervals measured by real clocks of reference. But this
time-transformation is by no means necessary in many occasions. 
 
iii) Step 3.- The third step is the three dimensional coordinate
transformation given by (\ref{1.17}). This is a passive, innocuous, change
of names to refer to the world-lines of the congruence. Its physical
meaning importance comes from being the last step to implement the
Relativity principle by the invariance of the line-element (\ref{1.1})
under the Lorentz transformations. But in practice its role is at most a
mere simplifying convenience which is not necessary in many occasions. 

Let us consider an example to clarify the above remarks. 
Let $F^{\alpha\beta}(T,X^i)$ be  
a second rank skew-symmetric tensor 
describing an electromagnetic field in the Galilean frame of
reference $S_0$. The electric and magnetic fields are then:

\begin{equation}
\label {1.25}
E^i=F^i_{.0} \quad B^{ij}=F^{ij}
\end{equation}
Let us now accept that for any other Galilean frame of reference with
congruence ${\cal C}_1$ the electric $E^a$ and magnetic $B^{ab}$ fields be given by: 

\begin{equation}
\label {1.30}
E^a=F^\alpha_{.\beta} \theta^a_\alpha U^\beta, \quad
B^{ab}=F^{\alpha\beta}\theta^a_\alpha\theta^b_\beta
\end{equation}
where 
$U^\alpha$ is the unit tangent to the world-lines of the congruence
${\cal C}_1$:

\begin{equation}
\label {1.26}
U^0=\frac{1}{\sqrt{1-v^2}}, \quad U^i=\frac{v^i}{\sqrt{1-v^2}}
\end{equation}
and where $\theta^a_\alpha$ are three arbitrary covariant vectors fields
orthogonal to $U^\alpha$.

Using the space coordinates transformation (\ref{1.6}) and keeping the
time coordinate $T$ unchanged leads to the following components:

\begin{equation}
\label {1.27}
u^0=\frac{1}{\sqrt{1-v^2}}, \quad u^i=0
\end{equation}

Three particular covariant vectors $N^\alpha$ orthogonal to $U^\alpha$
are:

\begin{equation}
\label {1.28}
N^a=dx^a: N^a_0= -v^a, \quad N^a_i=\delta^a_i \quad a,b,c,..=1,2,3
\end{equation}
From their definition we know that their components in the Galilean
system of reference $S_1$ are:

\begin{equation}
\label {1.29}
n^a_0=0, \quad n^a_i=\delta^a_i
\end{equation}

The same coordinate transformation leads to the following expressions for
the electromagnetic field in the Galilean frame of reference $S_1$:

\begin{eqnarray}
\label {1.21a}
f^{i0}&=&F^{i0} \\
\label {1.21b}
f^{ij}&=&F^{ij}+F^{i0}v^j-F^{j0}v^i
\end{eqnarray}
Each term in the r-h-s of these expressions is a function of $T$ and $X^i$ 
and the l-h-s are functions of $T$ and $x^i$ obtained using (\ref{1.6})
in the r-h-s.

In particular if we calculate the preceding
scalars (\ref{1.30}) in the system of reference $S_1$ with $\theta^a_i=n^a_i$ we
obtain:

\begin{equation}
\label {1.31}
e^a=f^\alpha_\beta n^a_\alpha u^\beta=\frac{1}{\sqrt{1-v^2}}f^a_{.0}, \quad
b^{ab}=f^{\alpha\beta} n^a_\alpha n^b_\beta=f^{ab}
\end{equation}

To obtain the result corresponding to a standard Lorentz
transformation we have to choose:

\begin{equation}
\label {1.32}
\theta^a=d\bar x^a
\end{equation}
where $\bar x^a$ have been defined in (\ref{1.17})

Notice that to transform vector or tensor components only the coordinate
transformations (\ref{1.6}) have been used. The time transformation
(\ref{1.14}) will be needed only if the original electromagnetic field
depends on $T$ and we want to use a time reference based on a clock
measuring proper-time along geodesics of ${\cal C}_1$. This may not be
convenient if the master clock on which a time distribution is based is
a clock which is at rest with respect to $S_0$. Neither is always the
most convenient choice to calculate the scalars (\ref{1.31}) using
(\ref{1.32}).

\section{Uniformly rotating frames of reference}
Let us consider again Minkowski's space-time with line element
(\ref{1.1}). It it will be now and then convenient to use cylindrical
space coordinates:

\begin{equation}
\label {2.4}
X=\rho \cos(\phi), \quad Y=\rho\sin(\phi), \quad Z=z
\end{equation}
in which case the line-element is:

\begin{equation}
\label {2.1}
ds^2=-dT^2+d\rho^2+\rho^2 d\Phi^2+dz^2
\end{equation} 

Let us consider the time-like congruence ${\cal C}_2$ defined by the following
parametric equations:

\begin{eqnarray}
\label {2.2a}
T&=&T \\
\label {2.2b}
X&=&x\cos(\omega T)-y\sin(\omega T) \\
\label {2.2c}
Y&=&x\sin(\omega T)+y\cos(\omega T) \\
\label {2.2d}
Z&=&z
\end{eqnarray}
where $x$, $y$ and $z$ are the initial conditions labelling each of
the world-lines of the congruence and $\omega$ is a constant. 
Using cylindrical coordinates the same congruence has parametric equations:

\begin{equation}
\label {2.3}
T=T, \quad \rho=\rho, \quad \phi=\varphi+\omega T, \quad z=z
\end{equation}

Considering the dual meaning of (\ref{2.2a})-(\ref{2.2d}), or
(\ref{2.3}), as a change of coordinates we get:

\begin{equation}
\label {2.5.0}
ds^2=-(1-\omega^2\rho^2)dT^2+2\omega(xdy-ydx)dT+dx^2+dy^2+dz^2
\end{equation}
which can be split as follows:

\begin{equation}
\label {2.5}
ds^2=-(\theta^0_2)^2+ds^2_2
\end{equation}
with:

\begin{eqnarray}
\label {2.6a}
\theta^0_2&=&-\sqrt{1-\omega^2\rho^2}dT+
\frac{\omega(xdy-ydx)}{\sqrt{1-\omega^2\rho^2}} \\
\label {2.6b}
d\hat s^2_2&=&dx^2+dy^2+dz^2
+\omega^2\frac{y^2dx^2+x^2dy^2-2xydxdy}{1-\omega^2\rho^2}
\end{eqnarray}
or:

\begin{equation}
\label{2.7.0}
ds^2=-(1-\omega^2\rho^2)dT^2+2\rho^2\omega d\varphi dT+d\rho^2+\rho^2 d\varphi^2+dz^2
\end{equation}
which can be split as follows:

\begin{eqnarray}
\label {2.7a}
\theta^0_2&=&-\sqrt{1-\omega^2\rho^2}dT+
\frac{\omega\rho^2 d\varphi}{\sqrt{1-\omega^2\rho^2}}\\
\label {2.7b}
d\hat s^2_2&=&d\rho^2+
\frac{\rho^2 d\varphi^2}{1-\omega^2\rho^2} +dz^2
\end{eqnarray}

The fact that (\ref{2.5.0}) does not depend on $T$ proves that the
congruence ${\cal C}_2$ is again a Killing congruence and therefore it has an
intrinsic meaning in the geometrical framework of Minkowski's
space-time. Each of its world-lines describes a circular motion with
constant angular velocity $\omega$ and thus ${\cal C}_2$ is the first
ingredient of a uniformly rotating frame of reference.

The second ingredient of a frame of reference must be a synchronization
associated to a foliation. To this end we choose any of the world-lines
of ${\cal C}_2$, say $L_0$, corresponding to initial conditions ($x_0$,
$y_0$, $z_0$) and define a foliation ${\cal F}_0$ as being the family of hyperplanes
orthogonal to $L_0$ and as associated synchronization the proper-time
scale along $L_0$. More precisely we define a new coordinate $t$ such that:

\begin{equation}
\label {2.8}
t-t_0=\int_0^{T_0}{\sqrt{1-\rho_0^2\frac{d\phi^2}{dT^2}}}\, dT=
\sqrt{1-\omega^2\rho_0^2}T_0
\end{equation}
where $\rho_0^2=x_0^2+y_0^2$, $t_0$ is an arbitrary origin of $t$ and 
$T_0(T,X,Y,Z;x_0,y_0,z_0)$ is the
value of $T$ at the intersection $E_0$ of $L_0$ with the hyperplane $\Pi$ of
the foliation ${\cal F}_0$ passing through the event with coordinates
$T,X,Y,Z$. Let $X_0$, $Y_0$  and $Z_0$ be the values of $X$, $Y$ and $Z$
at $E_0$, and let $\dot X_0$ $\dot Y_0$ and $\dot Z_0$ be the values of its derivatives
with respect to $T$. Then the equation of the hyperplane $\Pi$ is:

\begin{equation}
\label {2.9}
(X_0-X)\dot X_0+(Y_0-Y)\dot Y_0 +(Z_0-Z)\dot Z_0 -T_0+T=0
\end{equation} 
Taking into account that:
 
\begin{equation}
\label {2.10}
\dot X_0=-\omega Y_0, \quad \dot Y_0=\omega X_0, \quad \dot Z_0=0 
\end{equation}
Eq. (\ref{2.9}) simplifies to:

\begin{equation}
\label {2.11}
\omega(XY_0-YX_0) -T_0+T=0
\end{equation}
To find $T_0(T,X,Y,Z;x_0,y_0,z_0)$ this equation has to be solved keeping in mind that 
$X_0$ and $Y_0$ in:

\begin{eqnarray}
\label {2.11.1a}
X_0&=&x_0\cos(\omega T_0)-y_0\sin(\omega T_0) \\
\label {2.11.1b}
Y_0&=&x_0\sin(\omega T_0)+y_0\cos(\omega T_0) 
\end{eqnarray}
are functions of $x_0$, $y_0$ as well as the unknown $T_0$. And since this
equation is transcendental it has to be solved at some approximation. The
approximation that we consider below consists in using the Taylor
expansions of $X_0$, $Y_0$ and $T_0$ with respect to the variables $x_0$
and $y_0$ neglecting all monomials of order 3 or greater with
respect to $\omega$. 

If $x_0=y_0=0$ then $X_0=Y_0=0$ and we have:

\begin{equation}
\label {2.12}
T_0=T
\end{equation} 
Derivating (\ref{2.11}) with respect to $x_0$ we get:

\begin{equation}
\label {2.13}
\omega\left[X(\frac{\partial Y_0}{\partial x_0}+
\frac{\partial Y_0}{\partial T_0}\frac{\partial T_0}{\partial x_0})
-Y(\frac{\partial X_0}{\partial x_0}+
\frac{\partial X_0}{\partial T_0}\frac{\partial T_0}{\partial
x_0})\right]
-\frac{\partial T_0}{\partial x_0}=0
\end{equation}
or:

\begin{equation}
\label {2.14}
\frac{\partial T_0}{\partial x_0}=\omega[X\sin(\omega T)-Y\cos(\omega
T)]
\end{equation}
Similarly we get:

\begin{equation}
\label {2.15}
\frac{\partial T_0}{\partial y_0}=\omega[X\cos(\omega T)+Y\sin(\omega
T)]
\end{equation}
Equivalently, since from (\ref{2.2b}) and (\ref{2.2c}) we have:

\begin{eqnarray}
\label {2.15.1a}
x&=&\ X\cos(\omega T)+Y\sin(\omega T) \\
\label {2.15.1b}
y&=&-X\sin(\omega T)+Y\cos(\omega T) \\
\label {2.15.1c}
z&=&\ Z
\end{eqnarray}

we can write:

\begin{equation}
\label {2.16}
\frac{\partial T_0}{\partial x_0}=-\omega y, \quad
\frac{\partial T_0}{\partial y_0}= \omega x
\end{equation}
This process could be continued but the following derivatives would
be already of order $\omega^3$ and we shall stop it here.

$T_0$ is then at our approximation:

\begin{equation}
\label {2.17}
T_0=T+\omega x_0[X\sin(\omega T)-Y\cos(\omega T)]
+\omega y_0[X\cos(\omega T)+Y\sin(\omega T)] 
\end{equation}  
and therefore from (\ref{2.8}) we get at the corresponding
approximation:

\begin{equation}
\label {2.18}
t-t_0=(1-\frac12 \rho_0^2\omega^2)T+
\omega x_0[X\sin(\omega T)-Y\cos(\omega T)]
+\omega y_0[X\cos(\omega T)+Y\sin(\omega T)]
\end{equation}
This equation together with (\ref{2.15.1a})-
(\ref{2.15.1c})
completes the coordinate transformation from the Galilean frame of
reference $S_0$ to the uniformly rotating frame of reference $S_2$. 

From (\ref{2.18}), (\ref{2.15.1a}) and (\ref{2.15.1b}) we obtain:

\begin{equation}
\label {2.19}
T= (1+\frac12 \rho_0^2\omega^2)(t-t_0)+\omega x_0 y-\omega y_0 x
\end{equation}
which is the time-component transformation from the uniformly rotating
frame of reference $S_2$ to the Galilean frame of reference $S_0$.
Substituting $T$ from (\ref{2.19}) into (\ref{2.2b}) and (\ref{2.2c}) would yield the
space part of the transformation.

Up to this point we have completed two steps which are similar to those
of the preceding section:

i) Step 1 picked a rotating Killing congruence as the first ingredient
of a new frame of reference.

ii) Step 2 defined some convenient synchronizations.
But despite the similarities some relevant
differences with the pure Galilean case deserve to be mentioned explicitly: 

1) The congruences ${\cal C}_0$ and ${\cal C}_2$ share the bunch of
world-lines corresponding to the points of the axis of rotation but have
notorious well-known different intrinsic geometries. Moreover the domain
of ${\cal C}_2$ must be restricted to the domain $\omega\rho<1$ to keep
it time-like.

2) The synchronizations of $S_2$ depend on the world-line $L_0$ that defines the
scale of time, but those world-lines which are common to ${\cal C}_0$ and
${\cal C}_2$ are equally well adapted to both frames of reference.

There is another important qualitative difference between the case
considered here and that of the preceding section: namely that now the
quotient metric (\ref{2.7b}) is not Euclidean and therefore Step 3 there
does not make sense here because there are not Cartesian coordinates for
this metric. The far-reaching consequences of this fact which, as we are
told (\cite{Stachel}), played an important historical role in the genesis of General
relativity by A. Einstein, has been in our opinion under-estimated by
the relativity community ever since. First of all this means that rigid
bodies can not be compared in general if they are in different locations
or have different orientations. Another way of saying this is to say
that a rigid body can not be moved around. This shatters the very
foundations of metrology and therefore of physics. Similarly with the
concept of parallelism on which is based the idea that it makes sense
for two astronomers in two different locations to point their telescopes
in the same direction. 

In our opinion this unsatisfactory situation stems from a
misinterpretation of the line-element (\ref{2.7b}) as describing the
geometry of space in a rotating frame of reference. The point of view
that we develop below consists in defining the geometry of space by the
principal transform of (\ref{2.7b}), a concept that we introduced in
\cite{Bel96}, and in re-interpreting (\ref{2.7b}) as defining an optical
length, i.e. a length measured by a round trip transit time of light, instead of
a physical length, i.e. measured for instance with an stretched
ideally inextensible thread. 

The principal transform of (\ref{2.7b}) is by definition a metric with
line-element:

\begin{equation}
\label{2.20}
d\bar s^2=e^{2\mu}\left(d\rho^2+\frac{\rho^2}{1-\omega^2\rho^2}d\varphi^2\right)
+e^{2\nu}dz^2
\end{equation}
with $\mu$ and $\nu$ such that:

\begin{equation}
\label{2.21}
\bar R_{ijkl}=0
\end{equation}
and:

\begin{equation}
\label{2.22}
\hat g^{ij}\left(\hat\Gamma^k_{ij}-\bar\Gamma^k_{ij}\right)=0
\end{equation}
The first condition (\ref{2.21}) tells us that (\ref{2.20}) is Euclidean
and the second condition tells us that Cartesian coordinates of
(\ref{2.20}) are harmonic coordinates of (\ref{2.7b}). Both conditions
are necessary to make the association intrinsic and non ambiguous.

Requiring the function $\mu$ to be regular on the axis, the solutions for $\mu$ and 
$\nu$ of Eqs. (\ref{2.21}) and (\ref{2.22}) are:

\begin{equation}
\label {2.23}
\mu=\int^\rho_0\frac{du}{u}\left(\sqrt{1-\omega^2 u^2}-\frac{1}{1-
\omega^2 u^2}\right), \qquad \nu=0
\end{equation}

Step 3.- Now it makes sense to proceed with Step 3 requiring, if convenient,
the use of Cartesian coordinates of (\ref{2.20}). They are the following:

\begin{equation}
\label{2.23.1}
\bar x=\frac{e^\mu x}{\sqrt{1-\omega^2\rho^2}} , \quad 
\bar y=\frac{e^\mu y}{\sqrt{1-\omega^2\rho^2}}, \bar z=z
\end{equation}
so that (\ref{2.20}) becomes:

\begin{equation}
\label{2.23.2}
{d\bar s}^2={d\bar x}^2 +{d\bar y}^2 +{d\bar z}^2=
{d\bar\rho}^2+{\bar\rho}^2d\varphi^2+{d\bar z}^2
\end{equation}
with ${\bar\rho}^2={\bar x}^2 +{\bar y}^2$.
A system of orthonormal axes equally oriented all over the 
uniformly rotating frame of reference could now be obtained as 1-forms
differentiating (\ref{2.23.1}). 

Neglecting from now on all terms of order higher than $(\omega\rho)^2$ we have:

\begin{equation}
\label {2.24}
\mu\approx -\frac34 \omega^2\rho^2
\end{equation}

\begin{equation}
\label {2.26}
\bar\rho\approx \rho\left(1-\frac14 \omega^2\rho^2\right)
\end{equation}
and:

\begin{equation}
\label {2.27}
d\hat s^2\approx \left(1+\frac32 \omega^2\rho^2\right)(d\bar\rho^2+\bar\rho^2 
d\varphi^2)+dz^2
\end{equation}
or equivalently: 

\begin{equation}
\label {2.28}
d\hat s^2\equiv \frac{1}{c_1^2}d\bar\rho^2+\frac{1}{c_2^2}\bar\rho^2 
d\varphi^2+\frac{1}{c_3^2}dz^2
\end{equation}
where:

\begin{equation}
\label {2.29}
c_1=c_2\approx 1-\frac34 \omega^2\rho^2, \qquad c_3=1
\end{equation}

At this approximation the formulas (\ref{2.23.1}) and its inverse become:

\begin{equation}
\label {2.30.0}
\bar x=(1-\frac14\omega^2\rho^2)x, \quad
\bar y=(1-\frac14\omega^2\rho^2)y, \quad \bar z=z
\end{equation}
\begin{equation}
\label {2.30}
x=(1+\frac14\omega^2\bar\rho^2)\bar x, \quad
y=(1+\frac14\omega^2\bar\rho^2)\bar y, \quad z=\bar z
\end{equation} 

We are going to check an intuitive belief that is often promoted to a
fundamental principle called the Principle of local Lorentz invariance.
According to it one assumes that if the relevant time interval and 
domain of space are small enough then, at any location of a rotating
frame of reference, the usual Lorentz transformations can be used
ignoring every thing else except the instantaneous velocity of the
location with respect to a Galilean frame of reference at rest with
respect to the axis of rotation. 

Let us calculate $dT$ using (\ref{2.19}), (\ref{2.30}) and:

\begin{equation}
\label {2.31}
x_0=(1+\frac14\omega^2\bar\rho_0^2)\bar x_0, \quad
y_0=(1+\frac14\omega^2\bar\rho_0^2)\bar y_0, \quad z_0=\bar z_0
\end{equation}

Then, keeping the approximation to order $\omega^2$, let us calculate  $dX$ 
and $dY$ using (\ref{2.2b})-(\ref{2.2d}). And finally choose a world-line of ${\cal C}_2$ with initial 
conditions:

\begin{equation}
\label {2.34}
y_0=0, \quad t_0=0 
\end{equation} 
and evaluate the result when:
 
\begin{equation}
\label {2.32}
\bar x=\bar x_0, \quad \bar y=\bar y_0, \quad  \bar z=\bar z_0
\end{equation}
The final result is the following:

\begin{eqnarray}
\label {2.33a}
dT&=&(1+\frac12\omega^2\bar x_0^2)dt+\omega \bar x_0d\bar y \\
\label {2.33b}
dX&=&(1+\frac34\omega^2 \bar x_0^2)d\bar x  \\
\label {2.33c}
dY&=&(1+\frac54\omega^2 \bar x_0^2)d\bar y+\omega x_0 d\bar t \\
\label {2.33d}
dZ&=&d\bar z
\end{eqnarray}

These infinitesimal transformations have to be compared, at the
appropriate approximation, with the
Lorentz transformations corresponding to two Galilean frames of
reference when one of them moves with respect to the other with
constant velocity $v=\omega x_0$ in the $Y$ direction:

\begin{equation}
\label{2.33.1}
dT=(1+\frac12 v^2)dt+vd\bar y, \quad dX=d\bar x, \quad dY=(1+\frac12
v^2)dy+vdt, \quad dZ=d\bar z
\end{equation}
A glance to (\ref{2.33b})and (\ref{2.33c}) shows that these transformations do not
coincide, this meaning that the so-called Principle of Local Lorentz
Invariance is not valid in the framework that we have described that
includes the third step which led to (\ref{2.30}), but it is acceptable
if one is willing to renounce to (\ref{2.30}) and accept instead as
meaningful the local infinitesimal change of space coordinates:

\begin{equation}
\label{2.33.2}
d\tilde x=dx, \quad d\tilde y=(1+\frac12\omega^2\rho^2)dy, \quad 
d\tilde z=dz
\end{equation}

On the other hand one
sees, neglecting terms of order $\omega^2$ that the Principle of local
Galilean invariance is always satisfied as it was obvious from the beginning.

\section{The Wilson and Wilson experiment}
Let ($F^{\alpha\beta},\ K^{\gamma\delta}$) be the Minkowski's description
of an electromagnetic field in a medium with electric permittivity
$\epsilon$ and magnetic permeability $\mu$. The physical interpretation of
this couple of 4-dimensional skew-symmetric tensors comes from the
following identifications, where $E^\alpha$ is the electric field, $B^\alpha$ is 
the magnetic induction,
$D^\alpha$ is the electric displacement and $H^\alpha$ is the magnetic
field:

\begin{eqnarray}
\label {3.0a}
E^\alpha=F^\alpha_{\ \beta}u^\beta, \quad B^\alpha=-{\tilde F}^\alpha_{\ \beta}u^\beta \\
\label {3.0b}
D^\alpha=K^\alpha_{\ \beta}u^\beta, \quad H^\alpha=-{\tilde K}^\alpha_{\ \beta}u^\beta
\end{eqnarray} 
$u^\alpha$ being the unit vector tangent to the congruence defining
the motion of the frame of reference and ${\tilde F}^{\alpha\beta}$ being the dual of   
$F^{\alpha\beta}$.
In a Galilean frame of
reference co-moving with the medium ($u^i=0$) these formulas translate as follows:

\begin{eqnarray}
\label {3.1a}
E^i&=&F^i_{\ 0}, \quad B_k=\frac12\delta_{ijk}F^{ij} \\
\label {3.1b}
D^i&=&K^i_{\ 0}, \quad H_k=\frac12\delta_{ijk}K^{ij}
\end{eqnarray} 
The remaining components being zero.

The constitutive equations are:

\begin{equation}
\label {3.2}
D^i=\epsilon E^i, \quad H_k=\frac{1}{\mu}B_k
\end{equation}
and we shall use units such that for vacuum $\epsilon_0=\mu_0=1$. The
matching conditions at the boundary of a neutral medium with vacuum are:

\begin{eqnarray}
\label {3.3a}
(D^i_{+}-D^i_{-})n_i&=&0, \quad \delta_{ijk}(E^i_{+}-E^i_{-})n^j=0 \\
\label {3.3b}
(B_i^{+}-B_i^{-})n^i&=&0, \quad  \delta^{ijk}(H_i^{+}-H_i^{-})n^j=0
\end{eqnarray}
where $n_i$ is the normal to the boundary and where a super or sub index $+$ 
will refer to vacuum and $-$ will refer to the dielectric medium. 

If the medium is rigid, is uniformly rotating with respect to a 
Galilean frame of reference 
and adapted coordinates to the co-moving frame of reference are used
then the identifications (\ref{3.1a}) and (\ref{3.1b}) do not correspond anymore to. 
(\ref{3.1a}) and (\ref{3.1b}). Let us assume
that cylindrical coordinates are used and therefore the line-element
of Minkowski's metric in a Galilean frame of reference is (\ref{2.1}),
and (\ref{2.7.0}) in the rotating one.
The appropriate identification is then given by the following formulas, invariant 
under arbitrary synchronizations, derived from
(\ref{3.0a}) and (\ref{3.0b}) :

\begin{eqnarray}
\label {3.4a}
e^i=\xi^{-1}f^i_{\ 0}=g_{0\alpha}f^{i\alpha}, \quad b_k=\frac12\sqrt{\hat g}\delta_{ijk}f^{ij} \\
\label {3.4b}
d^i=\xi^{-1}k^i_{\ 0}=g_{0\alpha}g^{i\alpha}, \quad h_k=\frac12\sqrt{\hat g}\delta_{ijk}k^{ij}
\end{eqnarray}  
where $f^{\alpha\beta},\ k^{\alpha\beta}$ are the images of
$F^{\alpha\beta},\ K^{\alpha\beta}$ by the congruence transformation
(\ref{2.3}); $\xi=\sqrt{-g_{00}}$; $\hat g$ is the determinant of the
3-dimensional metric (\ref{2.7b})and $g_{0\alpha}$ are the corresponding
coefficients of(\ref{2.7.0}). Therefore:

\begin{equation}
\label {3.5.0}
\xi=\sqrt{1-\omega^2\rho^2}, \quad \hat g=\frac{\rho}{\sqrt{1-\omega^2\rho^2}}
\end{equation}

\begin{equation}
\label {3.5}
g_{00}=-(1-\omega^2\rho^2), \quad g_{11}=1, \quad g_{22}=\rho^2 \quad
g_{33}=1, \quad g_{02}=\omega\rho^2
\end{equation}   

The constitutive equations (\ref{3.2}) and matching conditions
(\ref{3.3a}) and (\ref{3.3b}) remain unchanged in form but they hold now for the
transformed fields:

\begin{equation}
\label {3.6}
d^i=\epsilon e^i, \quad h_k=\frac{1}{\mu}b_k
\end{equation} 
and:

\begin{eqnarray}
\label {3.7a}
(b^i_{+}-b^i_{-})n_i=0, \quad \delta^{ijk}(e^i_{+}-e^i_{-})n_j=0 \\
\label {3.7b}
(d^i_{+}-d^i_{-})n_i=0, \quad \delta^{ijk}(h^i_{+}-h^i_{-})n_j=0
\end{eqnarray}

We are going to use the preceding considerations to discuss the
Wilson and Wilson experiment. In this experiment a hollow dielectric cylinder
is rotated with constant angular velocity $\omega$ in a uniform
and constant magnetic field $B$ parallel to the axis of rotation. Two
brushes fixed with respect to the laboratory rub the inner and outer
cylindrical surfaces of radius, say $\rho_1$ and $\rho_2$, and the
electric potential difference between them $\triangle V$ is measured.
The results obtained in the experiments agree quite well with the
approximate formula which one obtains neglecting terms of order
$\omega^2\rho^2$ or smaller:

\begin{equation}
\label {3.8}
\triangle V=\frac12\mu
B\omega(1-\frac{1}{\epsilon\mu})(\rho_2^2-\rho_1^2)
\end{equation}
For our purposes though it is interesting to consider the following 
fully relativistic formula:

\begin{equation}
\label {3.9}
\triangle V=-\frac{\mu B}{2\omega}(1-\frac{1}{\epsilon\mu})
\ln\frac{1-\omega^2\rho_2^2}{1-\omega^2\rho_1^2} 
\end{equation}  

This formula can be derived using a variety of methods (\cite{Wilson}, \cite{Ridgely}, 
\cite{Klauber1}).
Our goal below is to show that (\ref{3.9}) can be derived from
(\ref{2.3}) and the line-element(\ref{2.7.0}), without using any non trivial time
transformation nor any redefinition of the space coordinates, thus
demonstrating that these two simple ingredients (\ref{2.3}) and
(\ref{2.7.0}) are all that it takes in some cases to implement Special
relativity physics. This explains why so many methods lead to the
correct result.

The tensor $F^{\alpha\beta}$ in Wilson and Wilson's experiment has a single non
zero component outside the cylinder, namely in cylindrical coordinates:

\begin{equation}
\label {3.10}
B^{+}_3=\rho F^{12}_{+}=B
\end{equation}
where $B$ is the uniform magnetic field. With the coordinate
transformation (\ref{2.3}) this contravariant field remains unchanged:

\begin{equation}
\label {3.10.0}
f^{12}_{+}=F^{12}_{+}
\end{equation}
so that:

\begin{equation}
\label {3.11}
b^{+}_3=\frac{\rho}{\sqrt{1-\omega^2\rho^2}} f^{12}_{+}=
\frac{B}{\sqrt{1-\omega^2\rho^2}} 
\end{equation}
but the line-element is now (\ref{2.7.0}) and therefore from (\ref{3.4a})
and (\ref{3.5}) we get:

\begin{equation}
\label {3.12}
e^1_{+}=\frac{\omega\rho B}{\sqrt{1-\omega^2\rho^2}}
\end{equation}
the remaining components being zero.

The non trivial matching conditions (\ref{3.7a}) (\ref{3.7b}) are:

\begin{equation}
\label {3.13}
d^1_{+}=d^1_{-}, \quad h^3_{+}=h^3_{-}
\end{equation}
wherefrom we get:

\begin{equation}
\label {3.14}
e^1_{-}=\frac{\omega\rho B}{\epsilon\sqrt{1-\omega^2\rho^2}}, \quad 
b^{-}_3=\frac{\mu B}{\sqrt{1-\omega^2\rho^2}}
\end{equation}
and therefore:

\begin{equation}
\label {3.14.0}
f^1_{-0}=\frac{\omega\rho B}{\epsilon}, \quad
f^{12}_{-}=\frac{\mu B}{\rho}
\end{equation}
Finally from:

\begin{equation}
\label {3.15}
f^1_{-0}=g_{00}f^{10}_{-}+g_{02}f^{12}_{-}
\end{equation}
we have:

\begin{equation}
\label {3.16}
f^{10}_{-}=\frac{1}{g_{00}}(f^1_{-0}-g_{02}f^{12}_{-})
\end{equation} 
or:

\begin{equation}
\label {3.17}
f^{10}_{-}=\frac{\mu\omega\rho B}{1-\omega^2\rho^2}
(1-\frac{1}{\epsilon\mu})
\end{equation}
Transforming back to the Galilean frame of reference to take into
account the fact that the brushes do not move we
obtain:

\begin{equation}
\label {3.18}
E^1=F^1_{-0}=-F^{10}_{-}=-f^{10}_{-}
\end{equation}
Using (\ref{3.17}) and 

\begin{equation}
\label {3.20}
E^1=-\partial_\rho V
\end{equation}
we obtain by a simple integration the formula (\ref{3.9})

\section{The Michelson-Morley experiment}

At the end of Sect. 2 we proposed to deal with the unsatisfactory
situation to which it leads the fact that the space metric (\ref{2.7b})
is not Euclidian by denying to it the role of describing the geometry
of space and to attribute this role to its principal transform (\ref{2.20}). This
raises the following question: what is then the meaning of
(\ref{2.7b})? The answer that we favor is that this metric describes a
crystal-like structure of vacuum that is responsible for an anisotropy
of the round trip speed of light coming from a distinction between
optical length, which is defined using (\ref{2.7b}), and geometrical
(or mechanical) length, which is defined using (\ref{2.20}). In other
words we propose to predict that the round trip velocity of light
$v_\gamma$ propagating in a direction $\gamma^i$ in a location with
coordinates $\bar x^k(\bar x,\bar y,\bar z)$ will be given by:

\begin{equation}
\label {4.1}
v_\gamma(\bar x^k)= \frac{\sqrt{\bar g_{ij}(\bar x^k)\gamma^i\gamma^j}}
{\sqrt{\hat
g_{ln}(\bar x^k)\gamma^l\gamma^n}}
\end{equation}   
With this interpretation it follows
that the unit vectors defined covariantly by the 1-forms:

\begin{equation}
\label {4.2}
\bar\theta^1=d\bar\rho, \quad \bar\theta^2=\bar\rho d\varphi, \quad
\bar\theta^3=d\bar z
\end{equation}
define the principal directions of the anisotropy, and that the scalars
(\ref{2.29}) are the corresponding speeds. An statement that can be
summarized by the following formula:

\begin{equation}
\label {4.3}
\frac{1}{v_\gamma^2}=\frac{\gamma_1^2}{c_1^2}+\frac{\gamma_2^2}{c_2^2}+
\frac{\gamma_3^2}{c_3^2}
\end{equation}
$\gamma^i$ being the cosines of the direction of propagation with
respect to the principal directions (\ref{4.2}). 

This point of view has been used to predict a non null outcome for those
experiments of the Michelson-Morley type that rotate on the horizontal
plane whatever is used as an oriented rigid standard of length.

If the location is at a colatitude $\theta$ and the direction of propagation 
lies on a horizontal plane making an angle $A$ with the East
direction then:

\begin{equation}
\label {4.4}
\vec\gamma= -\cos\theta\sin A\,\vec e_1+\cos A\, \vec e_2+\sin\theta\sin
A\,\vec e_3
\end{equation}
where $\vec e_i$ are the unit vectors corresponding to the principal
directions (\ref{4.2}), and we finally obtain:

\begin{equation}
\label {4.5}
v_\gamma\approx 1-\frac34 \omega^2R^2\sin^2\theta\left(1-\frac12
\sin^2\theta\right)-a_2\cos 2A
\end{equation}
where $R$ is the radius of the Earth, $\omega$ is its angular velocity and:

\begin{equation}
\label {4.6}
a_2=\frac38\omega^2R^2 \sin^4\theta
\end{equation} 

Most of the experiments of the Michelson-Morley type include to
improve its sensitivity a standard of length that rotates in the
horizontal plane. And therefore although the purpose of the
experiment is not to measure the parameter $a_2$ in fact they measure
it as part of a raw result to be used to test what they claim would be
violations of Special relativity had the experiment give a clear cut
result. In the experiment of Brillet and Hall $a_2$ was measured to
be $2.1\ 10^{-13}$ but this result was cited as being spurious
without further comment. The predicted result calculated from
(\ref{4.6}) is $3.1\ 10^{-13}$. We believe that the work presented in
this paper, as well as a few others that have preceded it, justifies
that the result of Brillet and Hall be checked. A few recent
experiments have improved the sensitivity of Brillet and Hall but
unfortunately they do not include a rotating arm and therefore they are
insensitive to the value of $a_2$.

\section*{Acknowledgments}

I gratefully acknowledge the help provided by A. Hern\'andez and A.
Molina at different stages of the preparation of this manuscript as well
as the stimulating discussions that followed a careful reading of it
by J.\ M.\ M.\ Senovilla.  
I also gratefully acknowledge the position of
visiting professor to the UPV/EHU that I have been holding since October
2000.


\begin{thebibliography}{99}

\bibitem{Then} J.\ W.\ Then, Am.\ J.\ Phys. \textbf{30} (1962)


\bibitem{Trocheris} M.\ Trocheris, Phil. Mag., \textbf{40} (1949) 

\bibitem{Takeno} H.\ Takeno, Prog. Theor. Phys., \textbf{7} (1952)

\bibitem{Herrera} L.\ Herrera and A.\ Di Prisco, arXiv:gr-qc/0201023


\bibitem{Mashhoon1} B.\ Mashoon, Phys. Letters A, \textbf{145}, No. 4 (1990).
\bibitem{Mashhoon2} B.\ Mashoon, arXiv:gr-qc/0301065


\bibitem{Bel96} Ll.\ Bel, Gen.\ Rel.\ and Grav. \textbf{28}, No. 9 (1996). 

\bibitem{AguirreBel} J.\ M.\ Aguirregabiria and Ll.\ Bel, 
Gen.\ Rel.\ and Grav. \textbf{33}, No. 11 (2001)

\bibitem{Bel02} Ll.\ Bel, arXiv: gr-qc/0210057


\bibitem{Wilson} M.\ Wilson and H.\ A.\ Wilson, Proc. R. Soc. London
Ser. A \textbf{89} (1913)

\bibitem{Pellegrini1} G.\ N.\ Pellegrini and A.\ R.\ Swift, Am.\ J.\ Phys. 
\textbf{63}, 8, (1995) 

\bibitem{Weber} T.\ A.\ Weber, Am.\ J.\ Phys. \textbf{65}, 10, (1997)

\bibitem{Ridgely1} T.\ Ridgely, Am.\ J.\ Phys. \textbf{66}, 2, (1998) 

\bibitem{Ridgely2} T.\ Ridgely, Am.\ J.\ Phys. \textbf{67}, 5, (1999) 

\bibitem{Pellegrini2} G.\ N.\ Pellegrini and A.\ R.\ Swift, Am.\ J.\ Phys. 
\textbf{65}, 9, (1997) 

\bibitem{Klauber1} R.\ D.\ Klauber, arXiv:gr-qc/0107035

\bibitem{Hertzberg} J.\ B.\ Hertzberg, S.\ R.\ Bickman, M.\ T.\ Hummon,
D.\ Krause, Jr., S.\ K.\ Peck and L.\ R.\ Hunter, Am.\ J.\ Phys. \textbf{69}, 6, (2001) 



\bibitem{MM} A.\ A.\ Michelson and E.\ W.\ Morley, Am.\ J.\ Sci. 
\textbf{34} (1887) 

\bibitem{Brillet} A.\ Brillet and J.\ L.\ Hall, Phys. Rev. Letters \textbf{42}, 9 (1979)

\bibitem{Muller} C.\ Braxmaier, H.\ M\"uller, O.\ Pradl, J.\ Mlynek, 
A.\ Peters and S.\ Schiller, Phys. Rev. Letters \textbf{88}, 1 (2002)

\bibitem{Wolf} P.\ Wolf, S.\ Bize, A.\ Clairon, A.\ N.\ Luiten, G.\ Santarelli, and 
M.\ E.\ Tobar, arXiv:gr-qc/0210049  

\bibitem{Lipa} J.\ A.\ Lipa, J.\ A.\ Nissen, S.\ Wang, D.\ A.\ Stricker
and D.\ Avaloff, to appear in Phys. Rev. Letters. 


\bibitem{Aspden} H.\ Aspden, Phys. Letters \textbf{85A}, 8 (1981)

\bibitem{Ching}  Ching-Chuan Su, EuroPhys. Lett., \textbf{56}, 2 (2001)

\bibitem{Klauber2} R.\ D.\ Klauber, arXiv:gr-qc/0103076, arXiv:gr-qc/0210106


\bibitem{ABMM} J.\ M.\ Aguirregabiria, Ll.\ Bel, J. Mart\'{\i}n and A.\
Molina, in {\it Recent Developments in Gravitation}, Eds. A.\ Feinstein
and J.\ Iba\~nez. World Scientific (1992)

\bibitem{BMM} Ll.\ Bel, J.\ Mart\'{\i}n and M.\ Molina, J. Phys. Soc.
Japan, \textbf{63}, 12 (1994)

\bibitem{BelMolina} Ll.\ Bel and A.\ Molina, Il Nuovo Cimento, \textbf{11B}, 6 (2000)
and arXiv:gr-qc/9806099 


\bibitem{Stachel} J.\ Stachel, in {\it Einstein and the History of General Relativity}, D.\ Howard, 
J.\ Stachel, Eds., {\it Einstein Studies}, \textbf{1} (1989)   

\end{thebibliography}
\end{document}